\documentclass[conference]{IEEEtran}
\usepackage{cite}
\usepackage{amsmath,amssymb,amsfonts}
\usepackage{algorithmic}
\usepackage{multirow,makecell}
\usepackage[ruled,vlined]{algorithm2e}
\usepackage{graphicx}
\usepackage{textcomp}
\usepackage{xcolor}
\usepackage[font=small,labelfont=bf]{caption}
\def\BibTeX{{\rm B\kern-.05em{\sc i\kern-.025em b}\kern-.08em
    T\kern-.1667em\lower.7ex\hbox{E}\kern-.125emX}}
\begin{document}

\title{Advancing Visual Specification of Code Requirements for Graphs\\}

\author{\IEEEauthorblockN{ Dewi Yokelson}
\IEEEauthorblockA{Computer and Information Science \\ University of Oregon \\ dewiy@cs.uoregon.edu}}

\maketitle

\begin{abstract}
Researchers in the humanities are among the many who are now exploring the world of big data.  They have begun to use programming languages like Python or R and their corresponding libraries to manipulate large data sets and discover brand new insights. One of the major hurdles that still exists is incorporating visualizations of this data into their projects. Visualization libraries can be difficult to learn how to use, even for those with formal training. Yet these visualizations are crucial for recognizing themes and communicating results to not only other researchers, but also the general public. This paper focuses on producing meaningful visualizations of data using machine learning. We allow the user to visually specify their code requirements in order to lower the barrier for humanities researchers to learn how to program visualizations. We use a hybrid model, combining a neural network and optical character recognition to generate the code to create the visualization. 
\end{abstract}

\section{Introduction}
\indent A new field has been formed within the traditional humanities fields called Digital Humanities (DH), or Computational Humanities. DH scholars are interested in harnessing the power of computers to analyze novels, poetry, and other writing to answer their research questions. Whether or not they have a specific question in mind, they want to manipulate their data and create visualizations to see if they can find surprising patterns. This indirect approach is called exploratory programming, and while it contains a lot of trial and error, it can yield fascinating results. \\
\indent Exploratory programming as a concept in the DH field has been developing over the past decade, where researchers use languages like Python or R and their corresponding libraries to provide new insight into their studies,  utilizing natural language processing (NLP) or other text analysis techniques. With the ability to now analyze higher volumes of text than was previously possible to do manually (i.e., distant reading vs. close reading), there has been an influx of new results. One of the pioneers of the exploratory programming approach is Nick Montfort; this approach is introduced in the text \textit{Exploratory Programming for the Arts and Humanities} \cite{b1}. Montfort discusses that when we are dealing with large amounts of data, the best way to both discover conclusions and communicate them is to create visualizations. Montfort points out that visualizations are crucial for exploratory programming in that they allow the researcher to see intermediary results and make decisions about what to look into next. Therefore, in many different regards, data visualizations are important for DH, and other fields. \\
\indent Most tools that are currently available for generating visualizations fall into one of two categories: (1) those that require significant programming experience and (2) those that never expose a user to the code. Researchers in non-Computer Science fields, and even some within it, don't necessarily have the programming background to make good use of the such toolkits in (1). Matplotlib, for example, requires an understanding of Python and object-oriented concepts to some degree. While new users can find code examples online and perhaps cobble something together, they are limited by what search terms they know, and by their knowledge of programming. Furthermore, official documentation for such libraries can be hard to find and confusing to those unaccustomed with how to decipher it. \\
\indent For the tools that require no coding (e.g., Tableau, Voyant), while they are made to be easy to learn and get started quickly, they ultimately limit user options as they don't provide access to the code directly. Users are left to either choose from the capabilities currently available, or request new features. Developers of these tools may take a long time to make and deploy these features, or never even implement them if they are not general enough. We wanted a tool that fits somewhere in between, that has a lower barrier to entry for those that are new to programming, but will allow users a full range of capabilities. This led to the idea that we could allow a user to visually specify what they want their graph to look like. This is what makes our tool novel compared to others. In this way it will also serve as an education aid, helping teach researchers who are simultaneously learning to code. \\
\indent Our primary target audience for this tool is DH researchers, who have started programming in Python environments, but still need help creating visualizations. We wanted our tool to generate fully executable example code that the researcher can then manipulate themselves. We trained a Convolutional Neural Network (CNN) to classify an input target image into one of many Matplotlib code templates. We also used optical character recognition (OCR) to scrape important text (e.g., title, labels) off of the example image and fill that into the template so that the researcher can have more insight into how the code works. If a user needs to generate a visualization for a paper or presentation, they would simply search online for a visualization that they would like to recreate. They would upload it to our tool and would get the Matplotlib code they need. The user would now only have to update some variable names and values to fill in their data and create their desired graph. This allows them to quickly have a working visualization but also grants them the ability to make as many modifications as they get more comfortable with programming. \\ 
\indent The tool we created is called \textit{Graph2Library}, or G2L, and was developed in collaboration with a working group of DH and Literature Researchers. Their feedback was instrumental in determining the output of G2L, as well as the types of visualizations G2L should be able to generate. We implemented it as a web-based tool so that it is easy to access and use. Contributions of this paper include: 
\begin{itemize}
\item A first step towards larger goal of Visual Requirements Engineering. Allowing researchers to specify their needs in terms of a graph visualization.
\item Creation of a hybrid model that combines a neural classifier with OCR to produce editable code.
\item Novel use of a code-based generator to produce a practically infinite synthetic training set.
\item Evaluation that showed users found the system to be effective and performance-enhancing.
\end{itemize}
We present recent related work and how our approach differs in Section II. In Section III, we discuss the approach and models. Further, Sections IV and V share our results and conclusions and Section VI discusses possible future extensions. 

\section{Related Work}
Related research work can be categorized into one of two areas: program synthesis (code generation), and creating hybrid models. 
\subsection{Program Synthesis}
\indent Perhaps the most similar and interesting research has been done in the area of program synthesis. One such project generated libmypaint and CAD (Computer-Aided Design) code from a target image to then reconstruct one that cannot be distinguished from the original\cite{b6}. In this project, they adversarially trained a recurrent neural network (RNN) to generate images and attempt to fool a discriminator CNN that was being trained to distinguish between the original and synthetic. They used Earth Mover's Distance (Wasserstein metric) to compare images and showed that it was more effective than $L^2$ Distance.\\
\indent The generative adversarial network (GAN) approach they took proved successful for generating their code, since they did not use a template based approach. However, we deemed this approach overly complicated for our use case of classifying graphs to their corresponding Matplotlib template. We like the GAN approach and may turn to it if we decide that we need more fine-grained code generation, i.e., lower-level Matplotlib code that is below our template level (see section VI Future Work). They worked with hand-drawn images, simple CAD models and, color portrait photographs. We will be using graphs, which on a complexity scale are somewhere in between the CAD models and color photographs, though because of the specific text on them we had to add another processing step. \\
\indent Other recent research in this area includes the use of RNNs for natural language to code, as Lin used \cite{b5}, which will be discussed in further detail in II-B Hybrid Models. Seq2seq is a popular approach for these RNNs, because it generates output vectors of a different size than the input vectors. This is necessary when the input, e.g., an English sentence, differs in size from the output, e.g., a code snippet. However, since we will be \textit{classifying} an image into one of a handful of code templates we do not need to use an RNN to generate variable length code output, but instead, a CNN.\\ 
\indent Another interesting foray into program synthesis is the creation of AutoPandas \cite{b19}. In this project they endeavor to solve a similar challenge, the complexity of modern libraries being too much for the novice programmer to learn. They take on the Pandas library, a Python based, Microsoft Excel-like data manipulating tool. They created a specification language of sorts by asking the user to provide an I/O example of how they want to transform their data. This example then gets passed to their neurally-backed program generator and code is generated and presented to the user. Their tool covers an impressive 119 transform functions in Pandas. \\ 
\indent There have also been projects that work in the opposite direction, generating visualizations from natural language\cite{b3} and from raw data\cite{b2}. We did not explore this avenue, as we liked the idea of simultaneously being able to teach the user how to code, instead of removing that visibility entirely.
\subsection{Hybrid Models}
\indent While much can be accomplished when using one machine learning model alone, there is a recent rise in using multiple techniques to fit the problem domain. Different than an ensemble model, where many algorithms work to classify one thing, a hybrid model is a conglomeration of two or more algorithms or technologies, to work on different parts on the problem. \\
\indent For example, Lin's project, Tellina, used an RNN and K-Nearest Neighbors (KNN). The RNN took natural language input (i.e., ``I want to delete all files from June 5") and generated a Bash code template that would execute this command. To deal with specific values, such as a date like June 5, they removed it from the input to the RNN and used KNN to determine it's variable type, which was then slotted back into the output code template \cite{b5}. Tellina is a web based tool that delivers results in real time. To evaluate their model, Lin conducted user studies that contrasted a control group of users that could only use the internet to get help completing some bash tasks with a group using Tellina. They summarized both quantitative and qualitative results from this study. Participants in the study spent an average of 22\% less time to complete tasks when using Tellina. They also found that users felt positively about Tellina helping them to complete tasks faster, and that it did not hinder them as much as it assisted them (results from a survey of the participants). We conduct a similar user evaluation, but take a slightly different approach as described in section IV. \\
\begin{figure*}
\includegraphics[width=\linewidth]{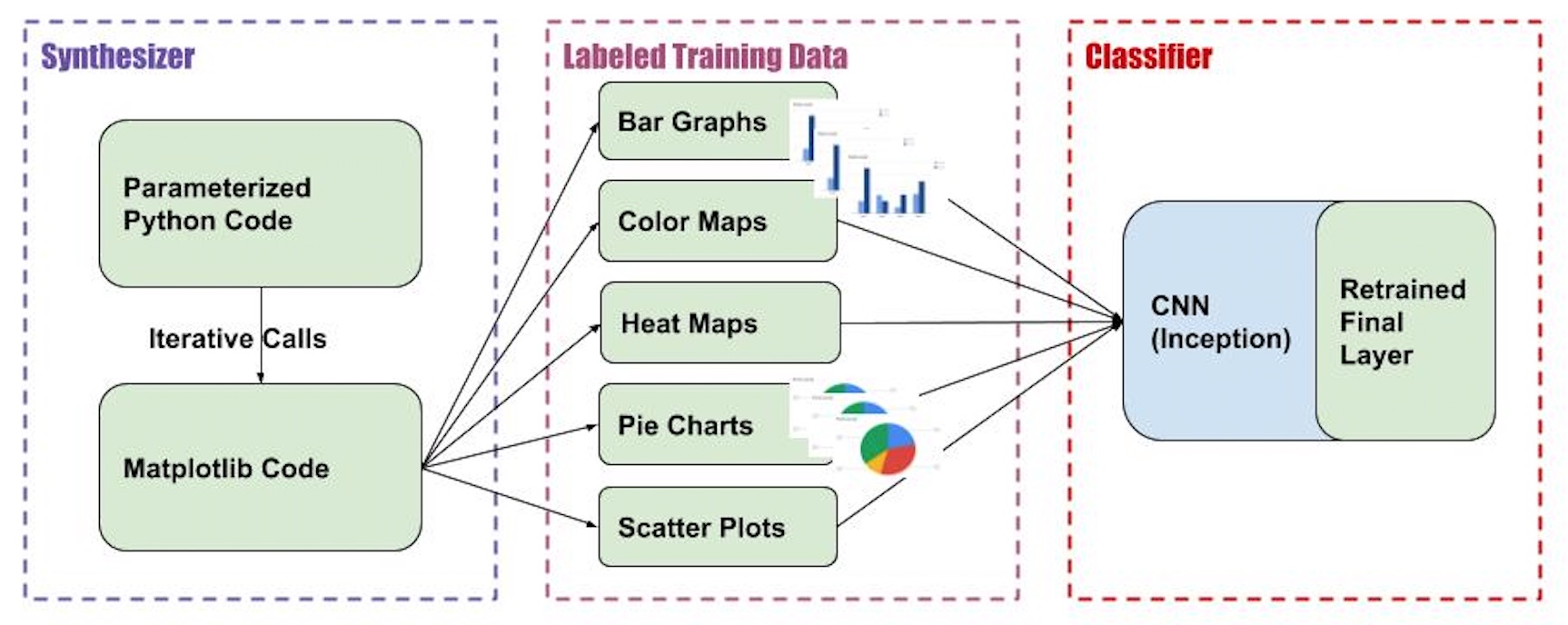}
\caption{Synthesized Training Data}
\end{figure*}
\indent Another project explored generating \LaTeX \space code from hand drawn images which met with some success, they supported primitive drawing commands to create a line, circle, and rectangle. They trained a neural network with attention mechanism to learn to infer specs which in turn can render an image in one or more steps. In addition to this neural architecture, they corrected mistakes by adding a Sequential Monte Carlo (SMC) sampling scheme which helps guide the output. Their results show that the combined model with the neural architecture and SMC significantly outperforms each on their own \cite{b4}. 

\section{Graph2Library}
\indent This section describes the approach for building G2L, to complete (A) training of the CNN, (B) OCR of the images, (C) putting the templates together, and (D) the final output. Figure 3 is provided for reference with corresponding subsections.
\subsection{Training the CNN}
\indent One challenge of using a neural network is that they typically require thousands of \textit{labeled} training examples in order to learn effectively. This labeled data can be difficult to acquire and a very manual process, especially when working with a new dataset. We navigated this issue in two ways, by (1) using transfer learning, and (2) by synthesizing hundreds of example graphs with their corresponding labels. 
\begin{table}[htbp]
\caption{Training Data Generated}
\begin{center}
\begin{tabular}{|c|c|}
\hline
\textbf{Graph Type}&\textbf{Number Generated} \\
\hline
Bar Graph & 300 \\
Scatter Plot & 300 \\
Pie Chart & 200 \\
Heat Map & 200 \\
Color Map & 200\\
\hline
\end{tabular}
\label{tab1}
\end{center}
\end{table}\\
\indent We chose to use Google's \textit{Inception} model as the pre-trained neural network for the transfer learning portion. This model has demonstrated 78.1\% accuracy on the ImageNet dataset \cite{b8}. We are able to use the trained model (on over a million images) and retrain only the final layer with our specific categories of graphs. We benefit from the earlier layers of the CNN already being able to recognize key image features, such as edges. Also, because we added hundreds of new training images of graphs specific to what we are classifying we achieved over 98\% training accuracy rate.\\
\indent To generate the specific training data, we wrote python code, that uses the Matplotlib library to generate graphs. We created six different classes of graphs: pie charts, bar graphs, stacked bar graphs, grouped bar graphs, scatter plots, and grouped scatter plots. We start with one code template per class, which is also considered the label of each synthesized graph. Each graph is generated, saved, and then the code template is automatically tweaked slightly (i.e., values, labels), while the bones of the template stay the same, and the process starts over again. See Figure 2 for an example of a synthesized grouped bar graph in which we manipulated the axes, data values, and title to generate many similar ones. 
\begin{figure}[h]
\centering
\includegraphics[width=\linewidth]{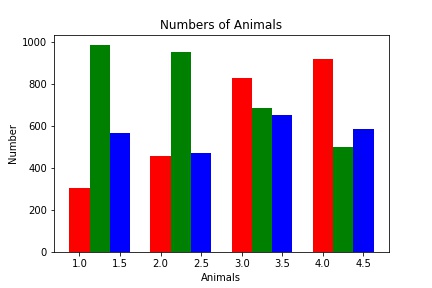}
\caption{Example of a synthesized training image}
\end{figure}
%\begin{figure}{0.98\linewidth}
%\includegraphics[width=\linewidth]{bar_graph_groups_97.jpg}
%\caption{Example of a synthesized training image}
%\end{figure}
\begin{figure*}
\includegraphics[width=\linewidth]{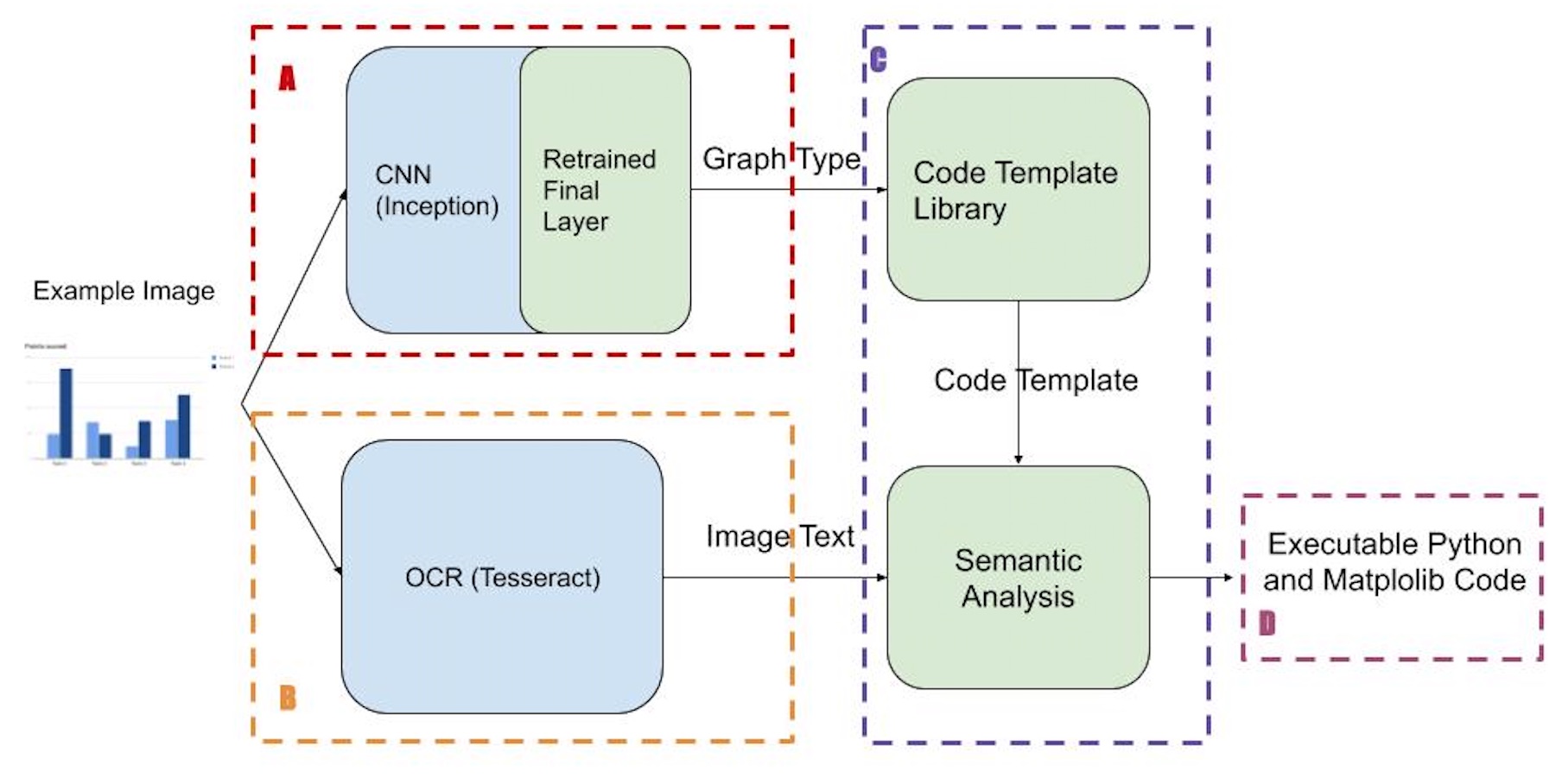}
\caption{\textbf{Figure 3. } }{\textbf{G2L Flow Diagram}}
\end{figure*}

\indent The end result of this part of the process is that we have a CNN that can classify our graphs into their types. Each type has its corresponding code template. Now with a test image we can classify it into a specific type of graph and the code template, once completed in the following step, will generate a similar graph, of the same type (see Figure 1 and section A of Figure 3).
\subsection{Optical Character Recognition}
\indent The neural net alone can classify an image to the correct graph type and corresponding code template but there are other elements of the graph that are important to preserve in order to generate fully executable code on the other end. Elements such as the number of labels on the x-axis of a graph can indicate how many data points are on the graph, or at the very least how many x-ticks are labeled. We use the Google library \textit{Tesseract} to conduct OCR on each test image. OCR allows us to gather the exact text, and information about its position on the image. We use the graph type obtained from the CNN and the text from our OCR step and can then perform more robust analysis to deliver a more complete code template. See Figure 3, section B for reference.\\
\indent The OCR step provides crucial information to G2L, specifically, any text on the image as well as where on the image it is located. This information is provided as the coordinates of the bounding box around each word found. Other helpful information, such as whether a set of words are considered part of a ``sentence" (based on position and proximity) is also part of what this step provides to us. During trials of \textit{Tesseract}, we discovered that while almost all horizontally aligned text is easily scraped, it is not able to scrape all text when it is rotated diagonally or vertically, but it can still pick up some. OCR is a tool that specializes in recognizing text, which means that it is not made for picking up lines, shapes, or other graphic features of the image. This is why we rely on the CNN to correctly categorize the image based on those features, and use OCR to get any additional information.

\subsection{Merging into Executable Code Templates}
\indent Once we know the type of the graph we are working with from the CNN, and have all the text scraped off of the image, we can conduct additional semantic analysis and customize the code template. This happens in section C of Figure 3. For example, a word or sequence of words on the far left of a bar graph is most likely to be a y-axis label. This differs from a pie chart in that if text is found in the same position on it, it is more likely to be a label of one section of the pie.  \textit{Tesseract} helps provide this information by denoting it a sentence or phrase.  Since graphs vary widely, these generalizations don't work in every scenario, but they provide a general guideline. We instead use default values such as ``Title" when our semantic analysis returns inconclusive results.\\
\indent There is also some analysis that can be conducted regardless of graph type. For example, if we find that our test image has a key or legend we know we will need to add the code for generating a key to the template. A title, while typically at the top and center of an image is also usually more than one word, see Algorithm 1 for how this is determined.\\

\begin{algorithm}
\SetAlgoLined
\KwResult{title}
 title = ``"\;
 \While{image has text}{
  get this\_word\;
  \If{this\_word in sentence}{ 
   \If{sentence position == top center}{
   return sentence;
   }
   }{
   \If{this\_word == (``vs" or ``v")}{
   word1 = word before this\_word;\\
   word2 = word after this\_word;\\
   return word1 + this\_word + word2;\\
   }{}{
   return ``Title"}\;
  }
 }
 \caption{Determining the Title of a Graph}
\end{algorithm}

\subsection{Final Output}
\indent The final output of our hybrid model is fully executable Python code that utilizes the Matplotlib library. See the following four images as an example of the process. Image 1 is an example of a graph that a user would find in a research paper that they want to recreate. This would be considered the target image fed into G2L and would be classified as a stacked bar graph. Once classified, OCR is conducted to scrape important features off of the graph and an output code template is produced. See Section D of Figure 3 for how this fits into the full G2L process.\\
\indent This code template contains specific data from the original image (e.g., Title, labels) so as to demonstrate to the user what parts of the graph are generated where. Also, variable names correspond with the object they represent, for example, where the data gets loaded in the template we use ``x = x\_data" so the user knows this has to do with the actual data.\\
\indent The user can then copy and paste the code template into their development environment of choice (e.g., Jupyter Notebook, Google Colab) and edit the code to include their own data. The output code template not only contains fully executable code, but also contains comments for how to update the code in some of the most common ways. This gives them the opportunity to customize in any way they need, but they will have working example code to start. The first thing they will want to change is to fill in their own data. We suggest the use of the Numpy or Pandas library to load data so as to keep solutions consistent and better assist the user. Figure 4 shows an example input image. Figure 5 shows the results of running the code template produced by Figure 4, a new graph that mimics Figure 4 but now has been updated to display the user's data.\\
\indent We heard from our working group that tools that require a lot of overheard for setup and getting started can be especially challenging. So when thinking about usage we decided to package all this functionality inside of a web application. This means it runs in the browser and requires no download or installation to begin using. A user simply uploads the file they want to get the code template for and they are given the output code template.

\section{Evaluation}
This section describes (A) the evaluation protocol, which was a two phased user study, (B) how we scored the accuracy of users, and (C) the post-evaluation survey results.

\begin{figure}[h]
\includegraphics[width=\linewidth]{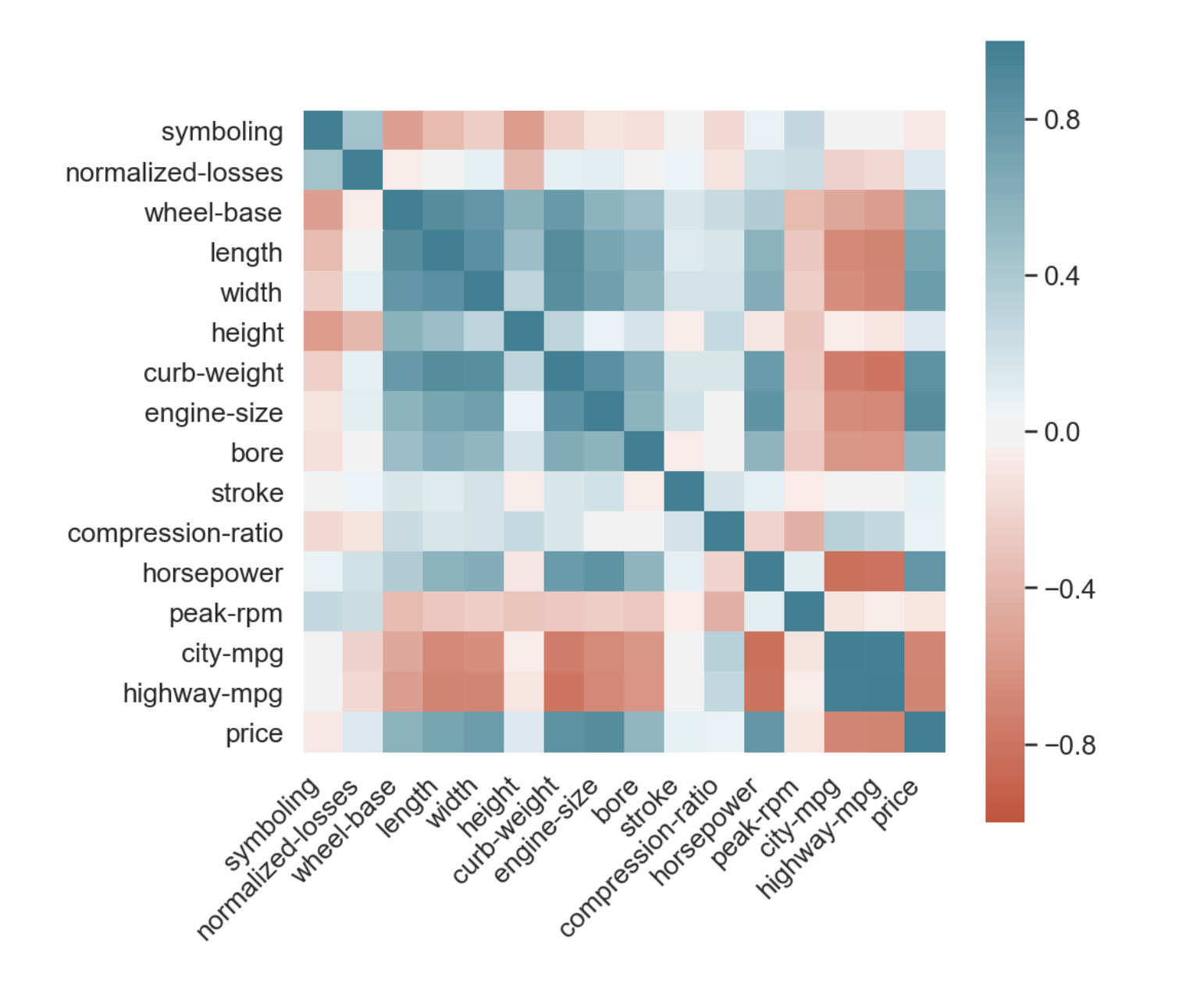}
\caption{Example Input or Target Graph from an External Source \cite{b7}}
\end{figure}

\begin{figure}[h]
\includegraphics[width=\linewidth]{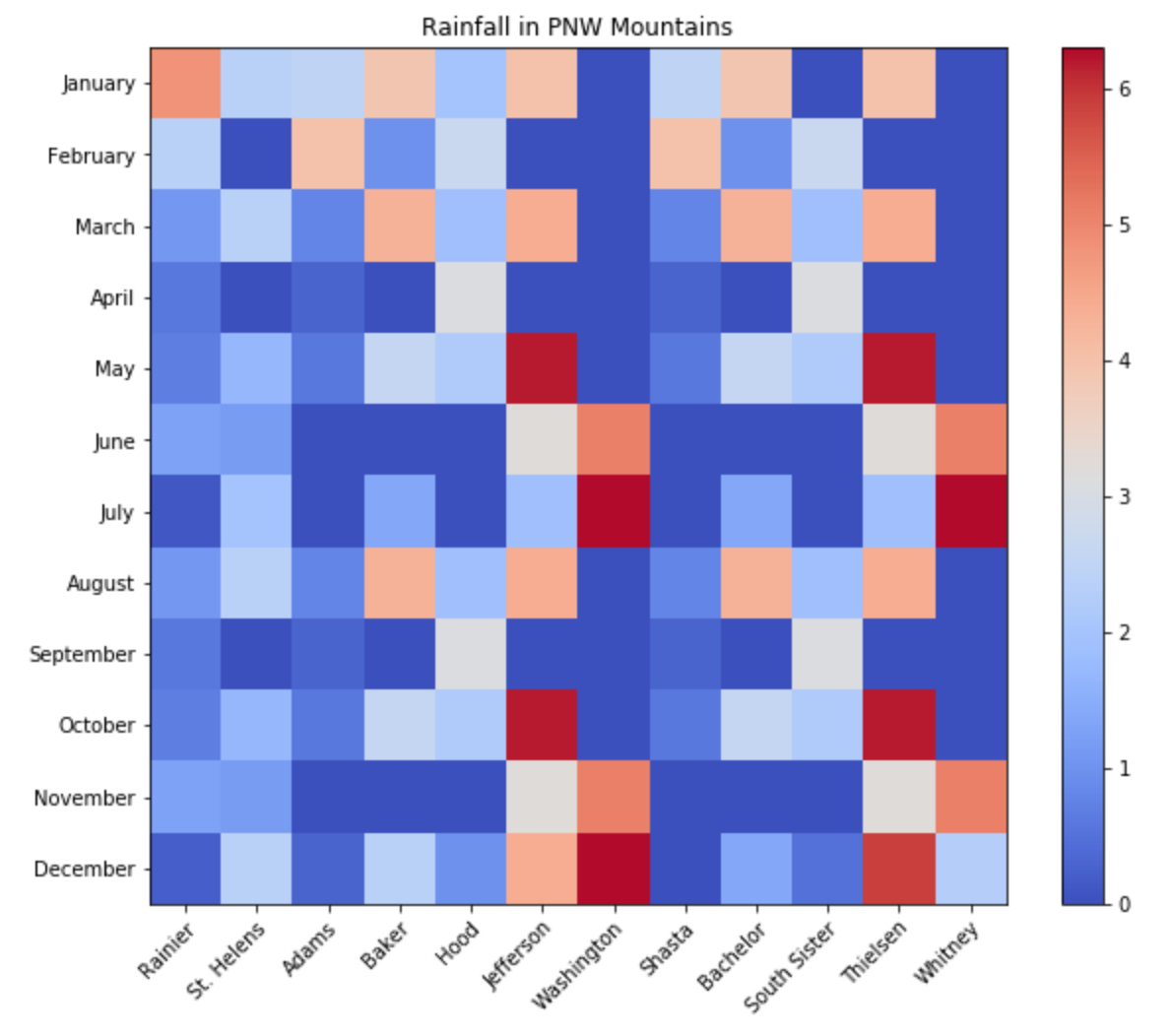}
\caption{Example of the Final Graph Generated from a Code Template Based on Figure 4}
\end{figure}

\subsection{Evaluation Protocol and Scoring}
\indent To evaluate our model we conducted two phases of user evaluation. The first was with three users that are studying Computer Science, and the second was with two Literature researchers. The first phase helped to work out any bugs in the evaluation process and get initial feedback on usability. The second phase was an evaluation on how G2L works for researchers in the humanities domain. In both phases, users were assigned the same series of tasks that contained milestones (T1-T6). In these tasks the user is provided with a starter Jupyter notebook that contains the necessary imports at the top, and pre-loaded data using the Numpy library. \\
\indent Selecting subjects was based on two main criteria: field of study and experience with specific technologies. For phase one we selected users who are undergraduate and graduate students in Computer Science. In phase two we selected graduate students in Literature (from the English Department). All users needed to have some experience with Python and Jupyter notebooks in order to complete the evaluation of G2L. Skills in Matplotlib and Pandas were not required. All self-reported skill levels for the four technologies mentioned are outlined in Table II. 

\begin{table}[htbp]
\caption{Pre-Survey Skill Self Rating}
\begin{center}
\begin{tabular}{|c|c|c|}
\hline
\textbf{Skill} & \textbf{\textit{User (Phase)}}& \textbf{\textit{Score$^{\mathrm{a}}$}} \\
\hline
\multirow{5}{*}{Python}& A (1) & 4\\
&B (1) & 5\\
&C (1) & 3\\
\cline{2-3} 
&D (2)& 3\\
&E (2)& 1\\
\hline
\multirow{5}{*}{Jupyter Notebooks}& A (1)  & 4\\
&B (1) & 4\\
&C (1) & 2\\
\cline{2-3} 
&D (2)& 2\\
&E (2)& 2\\
\hline
\multirow{5}{*}{Matplotlib}& A (1)  & 2\\
&B (1) & 4\\
&C (1) & 0\\
\cline{2-3} 
&D (2)& 1\\
&E (2)& 2\\
\hline
\multirow{5}{*}{Pandas}& A (1)  & 3.5 \\
&B (1) & 3\\
&C (1) & 2\\
\cline{2-3} 
&D (2)& 2\\
&E (2)& 1\\
\hline
\multicolumn{3}{l}{$^{\mathrm{a}}$Score on Scale 0-5.}
\end{tabular}
\label{tab1}
\end{center}
\end{table}
\indent To conduct the user evaluations we combined elements from the widely-used Thinking Aloud Protocol \cite{b9, b11, b13} with the measurement system in the Lemoncello et. al. paper \cite{b17}. Though the original Thinking Aloud Protocol process was developed over 30 years ago, it is still considered a tried and true method today for evaluating software for usability. Other research has shown that after evaluation by five users, this protocol finds 75-85\% of the problems in the software, and beyond this brings diminishing returns \cite{b10}. The process in general, involves a test user using the software in front of an observer. The test user vocalizes their actions and thoughts throughout the time it takes them to complete a task. The observer takes notes and the entire interaction is recorded and later transcribed and analyzed. We evaluated with five total users in two phases, the initial three and the final two. We recorded video and audio for all users and they were able to vocalize thoughts but also questions as they completed the tasks. Scripted answers were given to clarify the tasks if a user was unsure how to proceed. The time limit to complete all tasks in the evaluation was 30 minutes.\\
\indent Accuracy measurement of users completion of task milestones was based on the accuracy scale system that Lemoncello et. al. utilized in testing how well users could follow directions. This 0-6 scale adequately captured the different options for how well a subject completed tasks in the G2L evaluation. Audio and video was reviewed for each evaluation and each milestone was given a score on a 6 point scale which is defined as follows: 0 = unable; 1 = required intervention; 2 = asked for assistance; 3 = asked for verification; 4 = self-corrected; and 5 = correct and independent \cite{b17}. Some examples of this follow. If a user was unable to complete a task within the allotted time for the full evaluation then they received a 0 for that task milestone. If they made (a) an  initial erroneous attempt at the task before (b) reaching the milestone successfully and (c) without voicing any questions they then received an accuracy score of 4. These accuracy results are discussed further in the next section and the scores are encapsulated in Table III. 
\begin{table}[htbp]
\caption{Accuracy Score Per Task Milestone Per User}
\begin{center}
\begin{tabular}{|c|c|c|c|c|}
\hline
\multirow{2}{*}{\textbf{Task}} & \textbf{User} & \multicolumn{3}{|c|}{\textbf{Results}} \\
\cline{3-5} 
 &\textbf{(Phase)}& \textbf{\textit{Rating}} & \textit{\textbf{Average}} & \textit{\textbf{Std Dev}}\\
\hline
\multirow{5}{*}{1} & A (1)&  5 & \multirow{5}{*}{5} & \multirow{5}{*}{0}\\
& B (1)& 5 & &\\
& C (1)& 5  & &\\
\cline{2-3} 
& D (2)& 5  & &\\
& E (2)& 5  & & \\
\hline
\multirow{5}{*}{2} & A (1)& 5 & \multirow{5}{*}{5} & \multirow{5}{*}{0}\\
& B (1)& 5  & &\\
& C (1)& 5  & &\\
\cline{2-3} 
& D (2)& 5  & &\\
& E (2)& 5  & &\\
\hline
\multirow{5}{*}{3} & A (1)& 5 & \multirow{5}{*}{3.8} & \multirow{5}{*}{1.16}\\
& B (1)& 5  & &\\
& C (1)& 4  & &\\
\cline{2-3} 
& D (2)& 2  & &\\
& E (2)& 3  & & \\
\hline
\multirow{5}{*}{4} & A (1)& 5 & \multirow{5}{*}{5} & \multirow{5}{*}{0}\\
& B (1)& 5  & &\\
& C (1)& 5  & &\\
\cline{2-3} 
& D (2)& 5  & &\\
& E (2)& 5  & & \\
\hline
\multirow{5}{*}{5} & A (1)& 5 & \multirow{5}{*}{4.4} & \multirow{5}{*}{1.2}\\
& B (1)& 5  & &\\
& C (1)& 5  & &\\
\cline{2-3} 
& D (2)& 2  & &\\
& E (2)& 5  & &\\
\hline
\multirow{5}{*}{6} & A (1)& 5 & \multirow{5}{*}{4} & \multirow{5}{*}{2}\\
& B (1)& 5  & &\\
& C (1)& 5  & &\\
\cline{2-3} 
& D (2)& 5  & &\\
& E (2)& 0  & &\\
\hline
\end{tabular}
\label{tab1}
\end{center}
\end{table}

\indent In between phases one and two of the evaluation we made improvements to G2L based on feedback from phase one. The main changes were related to the content of the output code template. For example, all three users in phase one mentioned that they would like to be able to change the size of their new visualization more easily, even though this was not one of the assigned tasks. They felt it would be easier to see and also that it would be an update researchers would want to make to better fit presentations and papers. We then saw a user in phase two take advantage of the newly added functionality while completing their evaluation. Additional changes included: changing the phrasing for some of the comments, and adding more variables to the upper section of the template so that they were easier to update. 

\subsection{Evaluation Results and Analysis}
\indent The task set is outlined as follows. The users were given a Jupyter notebook that included  (1) an example heatmap to ``copy," (2) starter code (imports and pre-loaded data), and (3) instructions to recreate the graph using the Matplotlib code from G2L and the pre-loaded data. All users were told they could also use internet search at any time if needed. Successful completion of the task set was determined if users completed all six task milestones: (1) use G2L to get code template, (2) paste the code into Jupyter, (3) incorporate pre-loaded data into code template, (4) change the title, (5) remove the overlaid numbers on the heatmap, and (6) change the color schema of the visualization. The first four milestones are considered the ``core" milestones because they demonstrate the main function of G2L. Completion of these four milestones results in a visualization with the user's data. The latter two milestones are minor aesthetic updates.\\
\indent Five out of six users were able to complete all six milestones in around 15 minutes or less. One user had trouble with the final task milestone (6) and thus ran out of time, but was able to complete the other five. Total completion times for all milestones are outlined in Table IV, User E shows incomplete and 30 minutes because of task milestone six. As background, Milestone six was to change the color schema. In the template comments there was a link provided to official Matplotlib documentation that contained color options. User E did follow the link. Unfortunately, the documentation is a very verbose page, with the information needed towards the bottom. User E tried many of the code snippets at the beginning of the page, thoroughly read the unrelated information at the top, and never scrolled far enough. While more information could have been provided in the comment to guide the user, this is also a good example of when official documentation is too complicated to decipher. This is a common problem with official documentation; too often the provided examples are irrelevant, too simple, or too complex. \\
\indent Results show that there is a correlation between how a user rated themselves skill-wise and the speed at which they completed all tasks. Users A and B rated themselves significantly higher than other users in Python and Jupyter skills, and also slightly higher in Matplotlib and Pandas skills. Perhaps unsurprisingly they were able to complete the tasks the quickest, with both taking under 10 minutes for all six milestones. However, seeing as the target audience of the tool is less experienced users, it is encouraging to see that even those who rated themselves lower were able to complete the tasks in around 15 minutes. We would like to extend these same evaluations to more users to increase the sample size and thus the overall accuracy of our measurements.
\begin{table}[htbp]
\caption{Completion time for All Tasks}
\begin{center}
\begin{tabular}{|c|c|c|c|}
\hline
\textbf{User}&\multicolumn{3}{|c|}{\textbf{Results}} \\
\cline{2-4} 
 & \textit{\textbf{Phase}}& \textit{\textbf{Completed?}}& \textit{\textbf{Time (minutes)} }\\
\hline
A & 1& Yes & 9:30 \\
B & 1& Yes & 7:10 \\
C & 1& Yes & 15:50 \\
D & 2& Yes & 13:50 \\
E & 2& No & 30:00*\\
\hline
\multicolumn{3}{l}{*See Section 4B - Evaluation Results and Analysis}
\end{tabular}
\label{tab1}
\end{center}
\end{table}

\indent Analysis of the recorded sessions and observer notes show that G2L users were able to get the output code templates from G2L very quickly. Of the ``core" milestones, milestone three, manipulating the template to incorporate the pre-loaded data, proved the most challenging. With many hints incorporated into the code template, only two of the five users used internet search to help  complete the tasks. Only one of these searches was for a ``core" milestone, specifically milestone three. This shows that the core functionality of G2L was quite successful. 

\subsection{Post-Evaluation Survey Results}
\indent In addition to measuring time and accuracy scores for task completion, the user survey provided valuable insight. A total of six questions were asked of each user immediately after they completed their evaluation of G2L.  These questions and results are outlined in Table V. Results were generally positive, with three out of five users saying they would use G2L a lot in the future. Another important finding from the survey is that all users believed G2L helped limit the time they would spend searching the internet (average score 4.8). All five users felt that G2L helped them a lot overall. G2L helped the least in explaining the Matplotlib code. One remedy to this would be to update the comments and structure of the G2L output code template. 

\begin{table}[htbp]
\caption{Post-Evaluation Survey Scores}
\begin{center}
\begin{tabular}{|p{3cm}|c|c|c|c|}
\hline
\textbf{Skill} & \textbf{\textit{User}}& \textbf{\textit{Score$^{\mathrm{a}}$}} & Average & Std Dev\\
\hline
\multirow{5}{*}{\makecell{Would you use G2L\\ in the future?}}& A & 2 & \multirow{5}{*}{4} & \multirow{5}{*}{1.26}\\
&B& 3 & & \\
&C& 5 & & \\
&D& 5 & & \\
&E& 5 & & \\
\hline
\multirow{5}{*}{\makecell{How easy was it to\\ use G2L?}}& A & 4 & \multirow{5}{*}{4} & \multirow{5}{*}{0.63}\\
&B& 4 & & \\
&C& 5 & & \\
&D& 3 & & \\
&E& 4 & & \\
\hline
\multirow{5}{*}{\makecell{Did G2L limit the amount\\ of time you spent\\ searching the internet \\for code samples?}}& A & 5 & \multirow{5}{*}{4.8} & \multirow{5}{*}{0.40}\\
&B& 4 & & \\
&C& 5 & & \\
&D& 5 & & \\
&E& 5 & & \\
\hline
\multirow{5}{*}{\makecell{Did the comments \\within the code template \\ help you update your\\ visualization towards \\the targeted output?}}& A & 5 & \multirow{5}{*}{4.2} & \multirow{5}{*}{0.75} \\
&B& 4 & & \\
&C& 5 & & \\
&D& 3 & & \\
&E& 4 & & \\
\hline
\multirow{5}{*}{\makecell{Did G2L help you \\ understand the\\ Matplotlib code?}}& A & 5 & \multirow{5}{*}{3.8} & \multirow{5}{*}{1.17}\\
&B& 3 & & \\
&C& 5 & & \\
&D& 2 & & \\
&E& 4 & & \\
\hline
\multirow{5}{*}{\makecell{Did G2L help you\\ create your data \\ visualization overall?}}& A & 5 & \multirow{5}{*}{5} & \multirow{5}{*}{0}\\
&B& 5 & & \\
&C& 5 & & \\
&D& 5 & & \\
&E& 5 & & \\
\hline
\multicolumn{3}{l}{$^{\mathrm{a}}$Score on Scale 0-5.}
\end{tabular}
\label{tab1}
\end{center}
\end{table}

\section{Conclusion}
This paper describes the G2L web based tool and its demonstrated effectiveness in using a hybrid of machine learning, OCR, and rule-based text placement models to help researchers generate visualizations of their data. Using G2L means they are not required to have a detailed knowledge of Matplotlib, or which search terms they need to use on the internet to create a visualization. G2L requires some Python knowledge, but also provides assistance in learning how to use Python and Matplotlib. We have proven success with both Computer Science and Literature researchers using the tool and incorporating the results into a Jupyter notebook.\\
\indent While the evaluation in this project was done using Jupyter notebooks, it could have been any python interpreter. We chose Jupyter as it is one of the more commonly used interpreters, especially by researchers who want their work to be easy to recreate and verify. Visualizations display quite naturally within a Jupyter notebook alongside code and text, and rerunning code is easy, which make it a natural choice. However, there are also many pain points associated with Jupyter notebooks, including setup, kernel crashes, and code management \cite{b18}. Given these reasons may be enough to keep some away,  it would be interesting to conduct evaluations using different interpreters as well. The flexibility of the usage of the G2L output code template is a strength that makes it useful for not just researchers but a wide spectrum of programmers. \\
\indent In addition to helping DH researchers we have taken a step towards advancing the field of Visual Requirements Engineering. By allowing the users to specify what they want their graph to look like and generating code based on those specifications we have shown the success and potential of this approach and model. We have also proven that we can effectively synthesize the labelled training data that can often prove challenging to acquire for researchers using machine learning models. We have plans to make the G2L codebase publicly available and open source for people to view and add their own contributions, if desired. 

\section{Future Work}
\indent One direction for potential future work is to apply this same idea to visualizations using other code libraries (e.g., Plotly, Seaborn). To do so, we would need to abstract up from matplotlib. Fortuitously, there is related research being done on the grammar of graphics \cite{b16}. “A grammar of graphics is a framework which follows a layered approach to describe and construct visualizations or graphics in a structured manner.” \cite{b15} In fact, this extension, called a "layered grammar of graphics" was refined and used by the creators of \textit{ggplot2}, an R based visualization library \cite{b20}. Seven layered components of a graphic can be modeled as a pyramid, as identified by Sarkar in \cite{b15}. These components are data, aesthetics (i.e. axes), scale, geometric objects (i.e. bar), statistics (i.e. mean), facets (i.e. subplots), and coordinate system. The pyramid showing these seven components and how they build on each other is in Figure 6 below. This is promising in that it speaks to the potential of extending beyond just Matplotlib code to any type of visualization library. If these seven common building blocks can be identified by G2L, then they can be utilized to generalize the tool and generate different code output.

\begin{figure}[h]
\includegraphics[width=\linewidth]{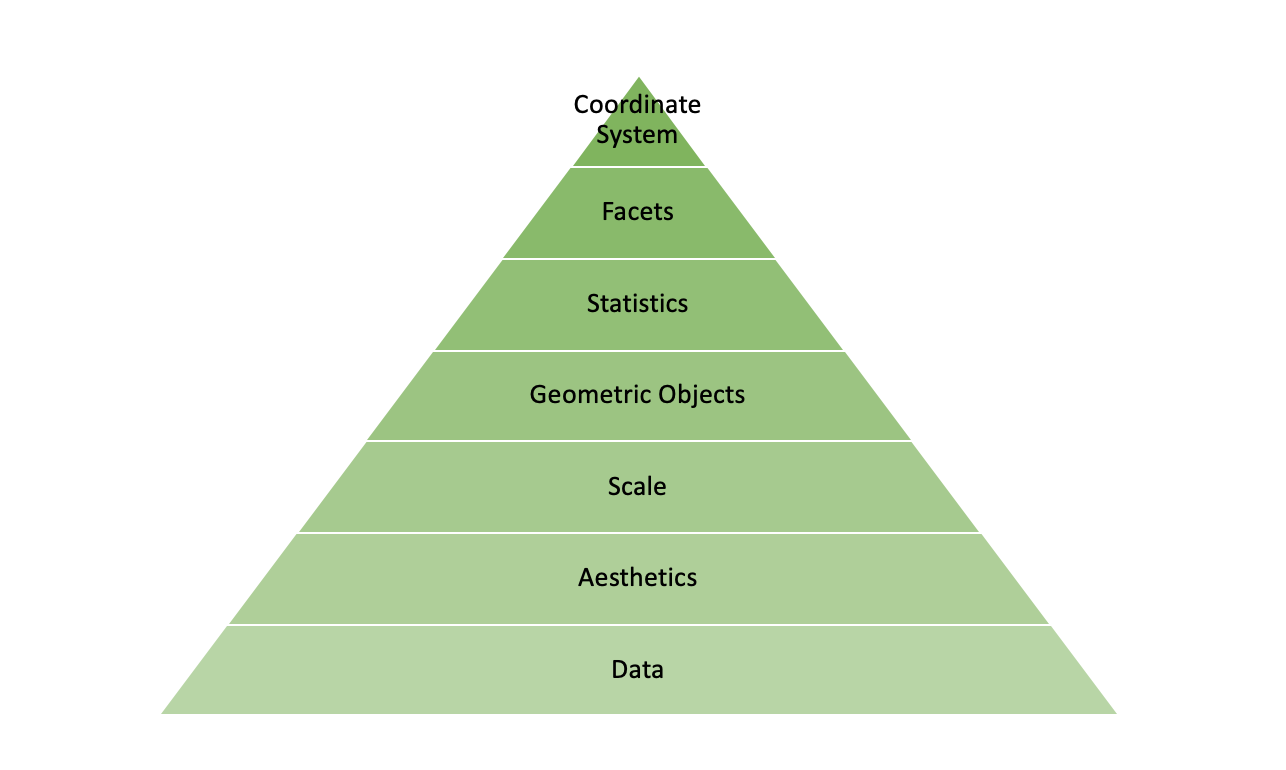}
\caption{Components of the Grammar of Graphics}
\end{figure}

\indent In an even broader extension of this work, visualizations could be treated as a new way to express software requirements in general. Visualizations have long been important in the Requirements Engineering field for comprehending traditional requirements. For instance, Magee et al. have built an animator for formal specifications using the executable language FSP \cite{b12}. We plan to explore whether animation (e.g., a video) can be used not only as an output once a specification has been developed, but also as the input itself to produce a specification. This is exciting since it extends single images into a sequence of images, and could make the programming of animations easier. Images are an extremely effective way to communicate ideas and results, and videos can be even more so, as they provide more dimensions of information. \\

\vspace{12pt}

\end{document}